\documentstyle{aipproc}
\input epsf.tex

\newcommand{\bm}[1]{\mbox{\boldmath $#1$}}

\begin{document}
\title{Statistical physics of adaptive correlation of agents in a market}

\author{David Sherrington, Juan P. Garrahan and Esteban Moro}
\address{Theoretical Physics, University of Oxford, 1 Keble Road,
Oxford, OX1 3NP, UK}

\maketitle

\begin{abstract}
Recent results and interpretations are presented for the thermal
minority game, concentrating on deriving and justifying the
fundamental stochastic differential equation for the microdynamics.
\end{abstract}

%\section*{Introduction}

Market economics poses several problems of potential interest and
challenge to statistical physics, involving the co-operative behaviour
of many agents whose actions involve mutual frustration and disorder,
both quenched and stochastic.  In a nutshell, speculators in an
idealized stock market are made up of buyers and sellers, each having
personal gain as their objectives, trying to buy low and sell high,
making their decisions based on commonly available information using
individual strategies, with their collective actions determining the
(time-varying) `right choices' and learning from experience.  From the
point of view of the market regulator, however, preference is for low
volatility and market efficiency. 

The {\it minority game} (MG) is a simple encapsulation of some of the
ingredients and issues of a market.  It consists of $N$ agents each of
whom at each step of a parallel dynamical process makes either of two
choices, with the objective of being in the minority overall.  The
agents have no direct knowledge of one another and make their
decisions based on purely global information $\vec{I}(t)$, available
equally to all.  Their decisions are determined through the
application to $\vec{I}(t)$ of individual strategy functions, each
agent having a small number of such strategies, drawn randomly and
independently from a large distribution at the outset and fixed
throughout the game.  At each time-step each agent employs (just) one
of his or her strategies.  Adaptation occurs through the development
of functions which determine their choices of strategy.

In the original formulation \cite{Challet97} the information 
$\vec{I}(t)$ was the minority choice over the last $m$ time-steps and the
adaptation was achieved through the cumulative award of points at each
time-step to the strategies which would have yielded the actual
minority choice at that step.  The strategy played by any agent at any
time-step was that of his/her strategies which currently had the
largest point-score.

A remarkable observation in simulations \cite{Savit99} was that the
variance in the minority choice became smaller than that of random
choice for large enough $m$, indicating correlation of the agents'
actions.  A critical memory length $m_c$ was observed for minimum
variance, with agents appearing to be frozen in their choices for
$m>m_c$, non-frozen for $m<m_c$. Moreover, it was shown that the
dependence on $m$ was through the scaling variable $D/N$, where
$D=2^m$ was the dimension of the space of strategies \cite{Savit99}.
Further simulations showed (i) these
results are unaffected by replacing the true history by a random 
$\vec{I}(t)$ \cite{Cavagna99}, indicating that as far as macroscopic
observables are concerned the `information' merely effectuates the
correlation; (ii) replacing the deterministic strategy-choices by
stochastic ones can significantly reduce the volatility for
information vectors of less than the critical length \cite{CGGS99}.

Here we consider the determination of a fundamental analytic theory
and report the derivation of the underlying stochastic differential
equation for the microdynamics \cite{Garrahan00}.  We concentrate on a
continuous formulation in which $\vec{I}(t)$ is a stochastically
randomly chosen unit-length vector on a $D$-dimensional hypersphere,
the strategies are quenched random vectors of length $\sqrt{D}$ in the
same space, $\vec{R}_i^\alpha, i=1, \ldots, N$ labeling the agents and
the $\alpha=1, \ldots, s$ their strategies.  The analogues of the
binary choices above are bids $b^\alpha_i(t)=\vec{R}^\alpha_i \cdot
\vec{I}(t)$.  The strategies which are actually used are indicated by
$\vec{R}^\ast_i(t)$.  The total bid at time $t$ is $A(t) =
\sum_i\vec{R}^\ast_i(t) \cdot \vec{I}(t)$. The point update rule is

\begin{equation}
P^\alpha_i(t+1)=P_i^\alpha(t) - b^\alpha_i(t)A(t)/N.
\end{equation}

For simplicity we specialize to $s=2$ and define
\begin{equation}
\vec{{\xi}}_i\equiv(\vec{R}^1_i - \vec{R}^2_i)/2, \ \ \vec{\omega}_i
\equiv(\vec{R}^1_i + \vec{R}^2_i)/2; \ \ p_i(t) = P^1_i(t) - P^2_i(t)
\end{equation}

In a generalized {\it thermal minority game} (TMG) the probability of
strategy use is

\begin{equation}
\pi_i^{1,2} (t) \equiv [1 + \exp (\mp \beta f (p_i(t)))]^{-1}
\end{equation}
and it is useful to define a `spin'
\begin{equation}
s_i(t)\equiv \pi^1_i(t) - \pi^2_i(t) = \tanh(\beta f(p_i(t))).
\end{equation}
In \cite{CGGS99} the choice $f(p)=p$ was employed, but here we
consider $f(p) = \rm{sgn}(p)\equiv z$ \cite{Garrahan00}.

We are interested in coarse-grained average behaviour on a time-scale
greater than the step-length in order to pass to a continuum-time
theory.  Equivalently, we take a time-scale $\Delta t$ with 
$\vec{I}(t)$ a differential random noise $\vec{I}(t) = \Delta\vec{W}(t)$ with
zero mean and variance $\Delta t$.  In the limit $\Delta t \to \infty$
a Kramers-Moyal expansion yields \cite{Garrahan00}

\begin{equation}
dp_i(t) = -(ND)^{-1} \vec{R}^\ast_i (t) \cdot \vec{\xi}_i dt + {\cal O}(dt^2)
\end{equation}
so that to ${\cal O}(dt)$ the information noise has been eliminated in favour
of an effective interaction between the agents and the averaged
variance becomes
\begin{equation}
\sigma^2\equiv N^{-1}\overline{\langle A(t)^2 \rangle} = 
(ND)^{-1}\sum_{ij}\langle
R^\ast_i(t) \cdot R^\ast_j(t)\rangle ,
\end{equation}
where the $\langle \cdot \rangle$ refer to a temporal average and the
bar to an average over the quenched disorder of the strategies.

At $T=0$ Eqs.\ (2) are deterministic and to leading order in $dt$ reduce to
\begin{equation}
d\bm{p}/dt = -\bm{\nabla}_{\bm{s}} {\cal{H}} |_{\bm{s}=\bm{z}}; \ \ \
\ \bm{p}\equiv (p_1,...p_N)
\end{equation}
\begin{equation}
{\cal{H}} = \sum_ih_is_i + {1\over 2} \sum_{i \neq j} J_{ij}s_is_j,
\end{equation}
\begin{equation}
h_i = (ND)^{-1}\sum_j\vec{\omega}_j \cdot \vec{\xi}_i, \ \
J_{ij}=(ND)^{-1}\vec{\xi}_i \cdot \vec{\xi}_j
\end{equation}

At finite temperature correlations between the fluctuations of the
right hand sides of Eqs.\ (5) are of the same order as the mean and
Eq.\ (7) must be replaced by a set of stochastic differential equations
\cite{Garrahan00}

\begin{equation}
d\bm{p}=-\bm{\nabla}_{\bm{s}} {\cal{H}} dt + {\cal{M}} \cdot d\bm{W}
\end{equation}
where ${\cal{M}} \equiv\{M_{ij}\}$ is the covariance matrix
\begin{equation}
M_{ij}[\bm{p}(t)]=\sum_{\kappa} J_{ik}J_{jk}[1-s^2_k(t)]
\end{equation}
and $\bm{W}(t)$ is an $N$-dimensional Wiener process of unit scale;
for $T=0 \ \ \ \ s_k(t)^2=1$ so the Wiener term has no weight.
Correspondingly, the Fokker-Planck equation for the probability
distribution of the $p_i$ is
\begin{equation}
\frac{\partial P}{\partial t} = -\sum_i \frac{\partial}{\partial
p_i}\left(\frac{\partial{\cal{H}}}{\partial s_i} P\right) + {1\over 2}
\sum_{ij}\frac{\partial^2}{\partial p_i\partial p_j} (M_{ij} P).
\end{equation}
The average volatility is given by 
\begin{equation}
\sigma^2\equiv N^{-1}\overline{\langle A(t)^2 \rangle} = 1 + 2
\overline{\langle{\cal{H}}\rangle}
\end{equation}

Eq.\ (10) is thus the fundamental microscopic equation from which the
macrodynamics should be calculable.  To check this we have compared
numerical evaluations of the volatility and the density of frozen
agents (those for whom $p_i(t)$ does not change sign after initial
transient) from Eqs.\ (7) and (10) with corresponding direct
simulations from Eqs.\ (1) and (3).  They are in perfect accord.  This
is shown explicitly in Fig. 1 for $T=0$.  Figs. 2a and 2b show the
effect of temperature as given by Eq.\ (10); direct simulation gave
results identical within statistical error.

\begin{figure}
\begin{center}
\leavevmode
\epsfxsize=3.0in
\epsffile{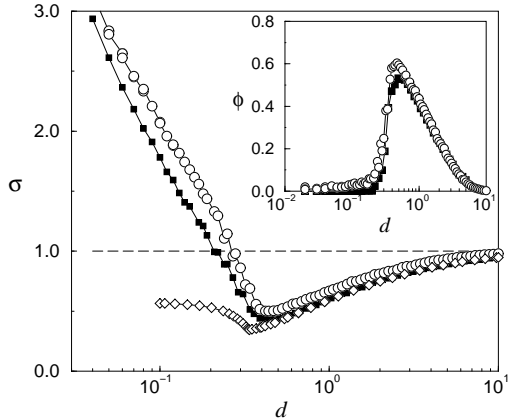}
\caption{Volatility $\sigma$ as a function of the reduced dimension
$d=D/N$.  Squares correspond to the original dynamics of Eq.\ (1),
circles to Eq.\ (7); $T=0$ and $p_i(0)\sim 0$.  Diamonds correspond to
minimization of $\overline{\cal{H}}$.  Inset: fraction of frozen
agents.}
\end{center}
\end{figure}

In \cite{CMZ00} $\overline{\langle{\cal{H}}\rangle}$ was evaluated for
$T=0$ on the assumption that the system equilibrated and hence was
equivalent to minimizing $\overline{\cal{H}}$.  The result is also
exhibited in Fig. 1 and can be seen to be good (and probably correct)
for $d = D/N >d_c$ but in error for $d<d_c$.  In fact, however,
Eq.\ (7) does not describe a simple descent dynamics since the
variables on the right and left hand sides are different and a metric
is needed to relate $p$ and $s$.  Substitution shows that the dynamics
in non-Markovian in $s$.  An explicit demonstration of
non-equilibration for $d<d_c$ follows from a simulation starting with
$|p_i(0)|\sim {\cal O}(1) \gg dt$, where $dt$ is the time-step.  This
is illustrated in Fig. 3a \cite{Garrahan00}.

\begin{figure}
\begin{center}
\leavevmode
\epsfxsize=3.0in
\epsffile{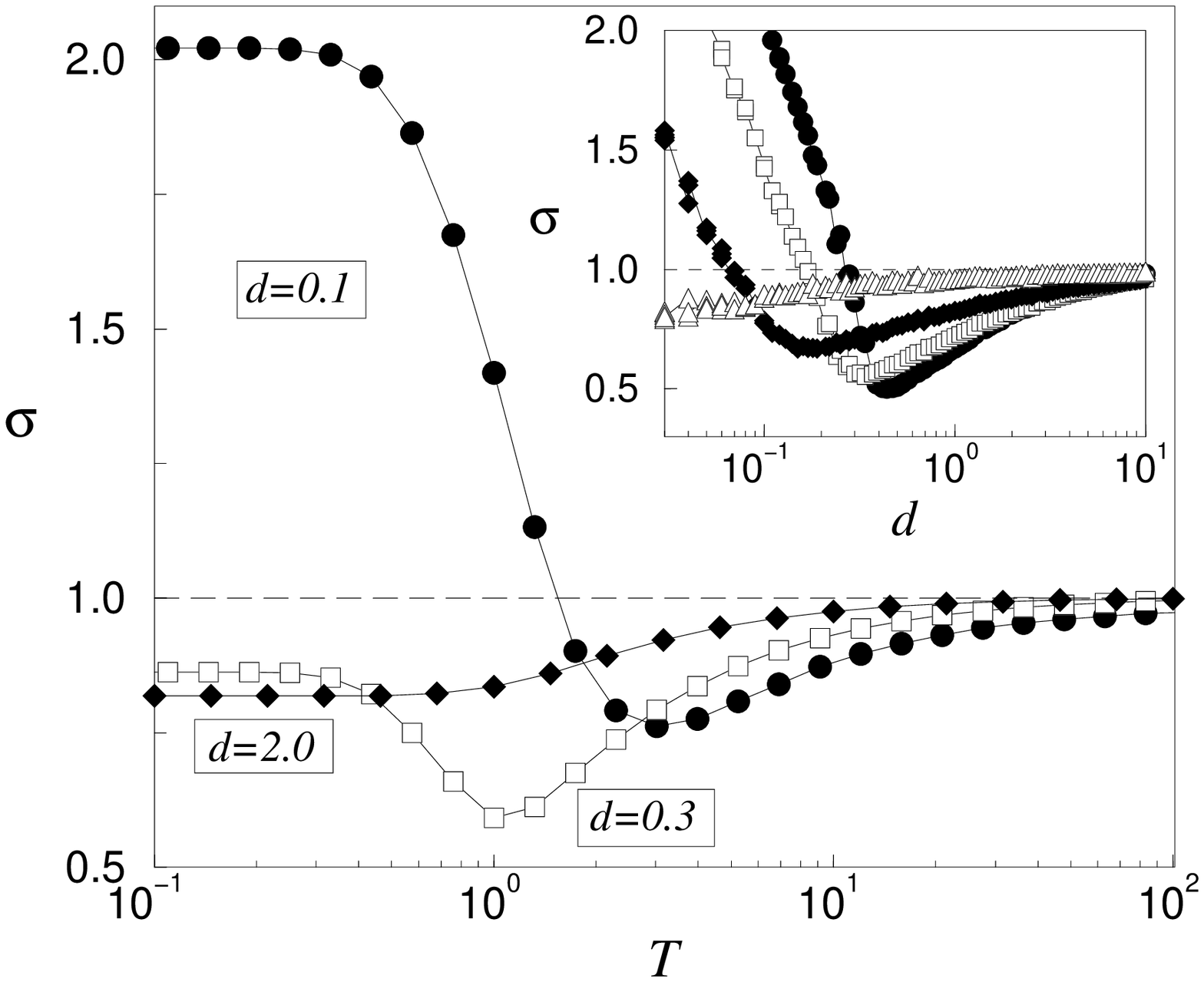}
\epsfxsize=2.8in
\epsffile{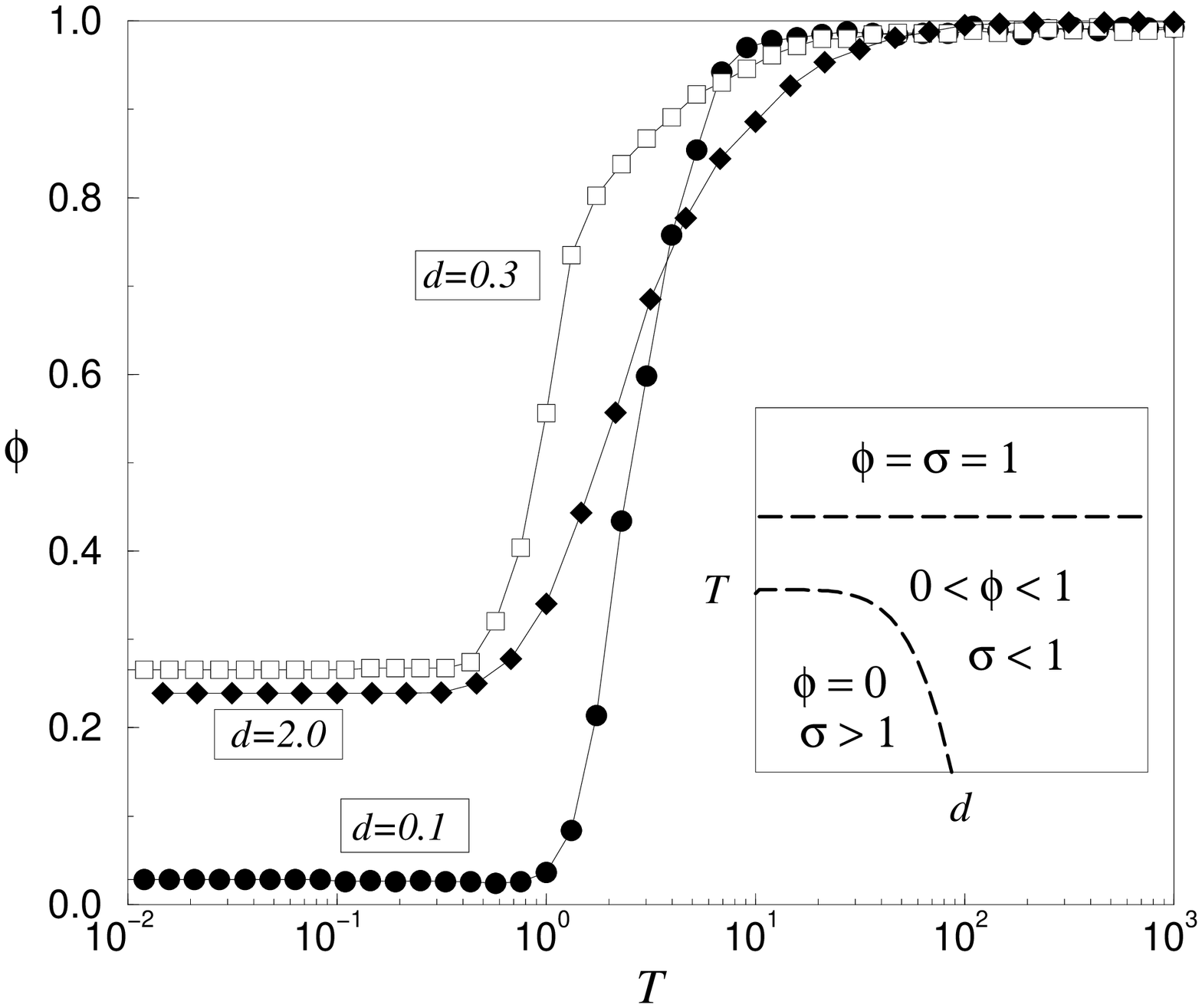}
\caption{(a) Volatility as a function of the temperature from the
continuous dynamics Eq.\ (10);$f(p)={\rm{sgn}}(p)$, $p_i(0)\sim 0$.
Inset: volatility as a function of $d$ for $T=10^{-3}$, $1$, $2$, and
$10$. (b) Fraction of frozen agents as a function of $T$.  Inset:
schematic cross-over phase diagram.}
\end{center}
\end{figure}

It is also tempting to compare with a Hopfield neural network which is
characterizable by an effective Hamiltonian
\begin{equation}
H=-{1\over 2}\sum_{i \neq j} J_{ij}^H\sigma_i\sigma_j; \ \ J_{ij}^H =
N^{-1}\sum_{\mu=1}^{p=\alpha N} \vec{\xi}^\mu_i \cdot \vec{\xi}^\mu_j
\end{equation}
where the $\{\xi^\mu_i\}$ are quenched random patterns; indeed it was
by the application of techniques devised for (14) that \cite{CMZ00}
minimized $\overline{{\cal{H}}}$.  Clearly there is a difference of
sign between Eqs.\ (8) and (14) but one might be tempted to anticipate
that this will merely suppress the retrieval attractors while
maintaining the spin glass state, in analogy with the SK model, and
then attribute the reduction in energy compared with the random state
to spin-glass binding.  However, this is false; the Hopfield
spin-glass solution is not symmetric under change of sign of $\beta$.
Rather, the random-field term of Eq.\ (8) is crucial in reducing the
ground state energy of ${\cal{H}}$ and the volatility below their
random-state values.  This is demonstrated explicitly in Fig. 3b which
shows the effect of choosing $\vec{R}^1_i$ randomly but $\vec{R}_i^2$ as
\begin{equation}
\vec{R}^2_i = -(1-\lambda) \vec{R}^1_i + \lambda \vec{\tilde{R}}_i
\end{equation}
where $\vec{\tilde{R}}_i$ is also chosen randomly \cite{CMZc00,prepa}.  For
$\lambda = 0, \vec{\omega}_i=0, h_i=0$ and the volatility never falls
below random, although again there appears a (different) critical
$d_c$ separating regimes (worse-than-random and random).

As noted, above we have used $f(p) = {\rm sgn} (p)$ in the
simulations.  If instead $f(p) = p$ is employed, then for $d>d_c$ the
system iterates over a long time to its zero-temperature behaviour
\cite{CMZb00} since the mean $|p_i(t)|$ grows quasi-continuously and
$s_i(t)$ saturates to its zero-temperature value, which being $\pm 1$
eliminates the effects of the second term of Eq.\ (10).  However, for
$d<d_c$ there continues to be an improvement with temperature, to an
optimal value which is better than random and is reached at a
temperature of ${\cal O}(1)$ \cite{CGGS00}, but without any further
rise to the random value; in this case $p_i(t)$ fluctuates around
$p_i(0)$.

Finally we remark on the relationship with the crowd-anticrowd concept
of \cite{HJJH00}, where a crowd is a group of agents playing the same
strategy and the corresponding anticrowd play the opposite strategy.
From Eq.\ (6)
\begin{equation}
\sigma^2=D^{-1}\sum_\mu\langle n_\mu(t)\rangle^2; \ \ n_\mu =
N^{-1/2}\sum_i\vec{R}^\ast_i(t) \cdot \vec{e}_\mu
\end{equation}
where $\vec{e}_\mu$ is a unit vector in the $\mu^{th}$ Cartesian
direction of the D-dimensional space.  $n_\mu$ then formalizes the
notion of the number of agents in crowd $\mu$ minus the number in the
corresponding anti-crowd.  The qualitative difference of $d < d_c$ and
$d > d_c$ then follows from the recognition that for $d\ll d_c$ the
vectors $\vec{R}^\alpha_i$ are densely distributed on the D-sphere
permitting $n_\nu \sim {\cal O}(N)$, while for $d\gg d_c$ they are sparsely
distributed so all $n_\nu \sim {\cal O}(1)$.

\section*{Acknowledgments}
 
We are grateful to Andrea Cavagna, Irene Giardina and Matteo Marsili
for useful comments and discussions.  We acknowledge financial support
from EPSRC Grant No.\ GR/M04426 and EC Grant No.\ ARG/B7-3011/94/27.

\begin{figure}
\begin{center}
\leavevmode
\epsfxsize=3.0in
\epsffile{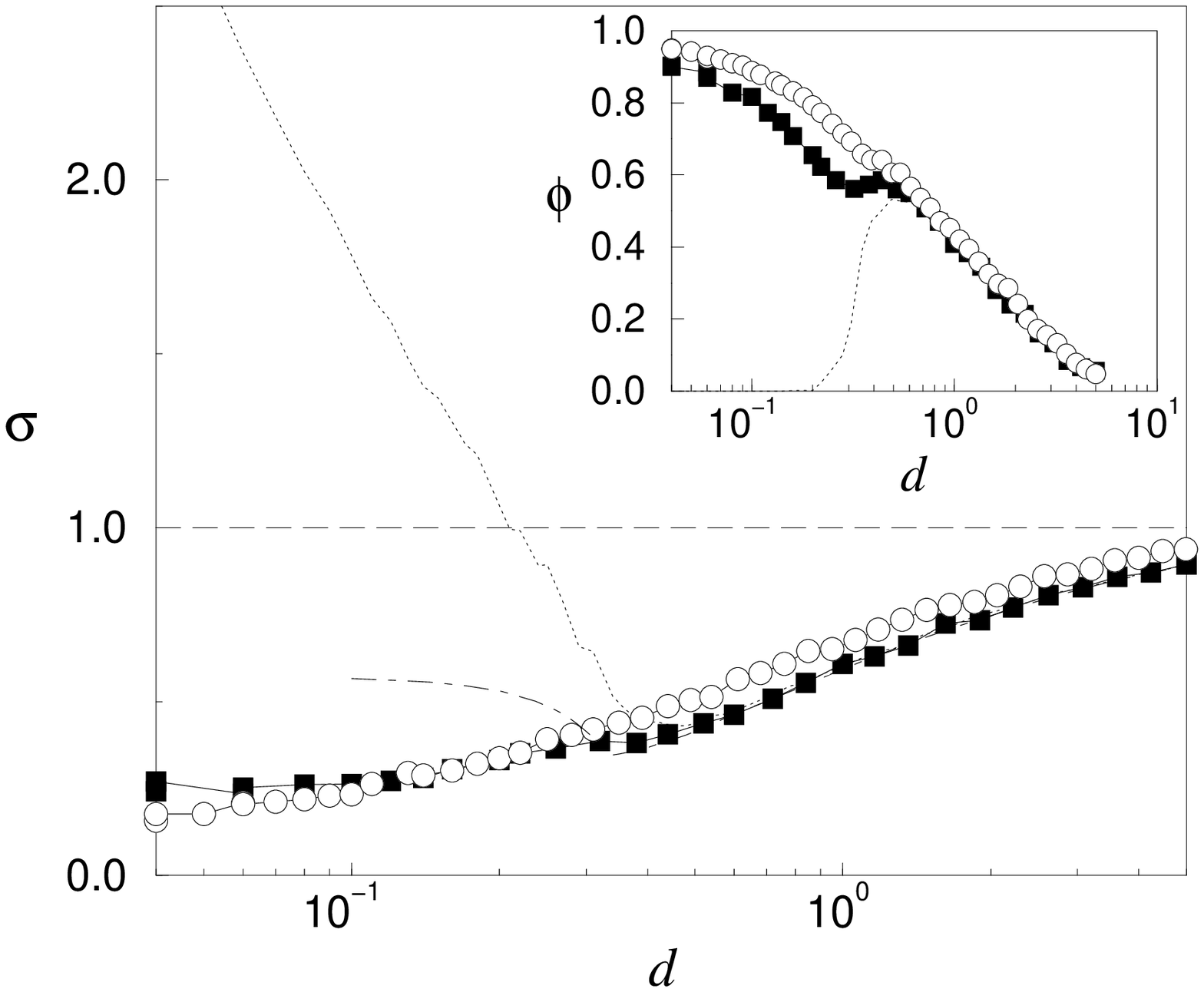}
\epsfxsize=2.8in
\epsffile{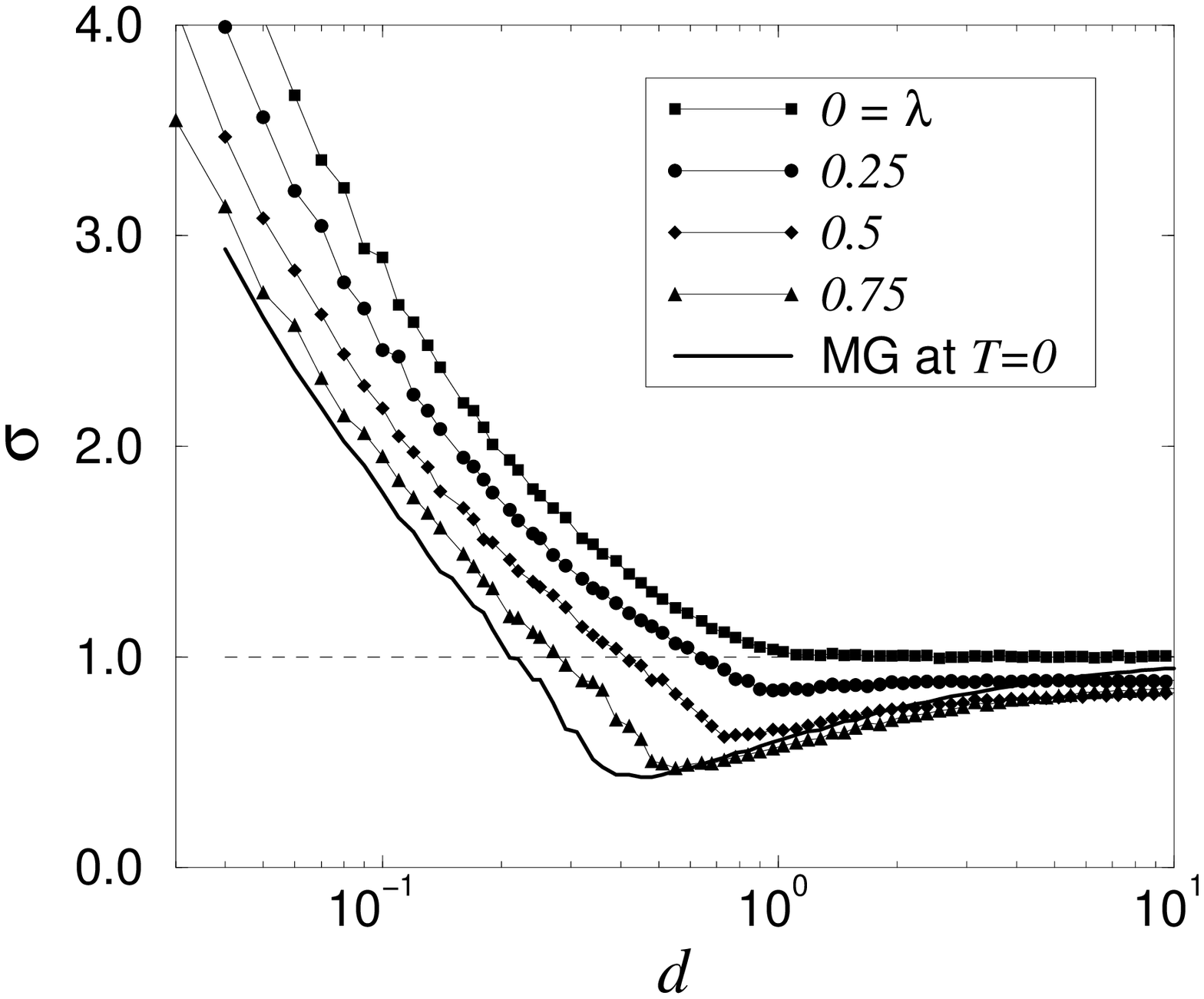}
\caption{(a) Volatility as a function of $d$ for random initial
conditions $|p_i(0)|\sim {\cal O}(1) \gg dt$ for original dynamics and from
Eq.\ (10).  (b) Volatility for partially anticorrelated
$\vec{R}_i^{1,2}$; $\vec{R}^2_i=-(1-\lambda)\vec{R}_i^1 + 
\lambda \vec{\tilde{R}}_i$; $\vec{R}_i^1, \vec{\tilde{R}}_i$ random.}
\end{center}
\end{figure}

\end{document}